\documentclass[twocolumn,preprintnumbers,prb,fleqn,superscriptaddress]{revtex4}
\usepackage{graphicx}

\begin{document}

\title{Thermal conductivity of graphene in Corbino membrane geometry}

\author{Clement Faugeras}
\affiliation{Laboratoire National des Champs Magn\'etiques Intenses,
CNRS, 25 rue des Martyrs, 38042 Grenoble, France}\email{clement.faugeras@lncmi.cnrs.fr}
\author{Blaise Faugeras}
\affiliation{Laboratoire J.-A. Dieudonn\'e, Universit\'e de Nice,
Parc Valrose, 06108 Nice cedex 2 France}

\author{Milan Orlita}
\affiliation{Laboratoire National des Champs Magn\'etiques
Intenses, CNRS, 25 rue des Martyrs, 38042 Grenoble, France}
\affiliation{Also at : Institute of Physics, Charles University,
Ke Karlovu 5, CZ-121 16 Praha 2, Czech Republic and Institute of
Physics, ASCR, Cukrovarnická 10, CZ-162 53 Praha 6, Czech
Republic.}
\author{M. Potemski}
\affiliation{Laboratoire National des Champs Magn\'etiques Intenses,
CNRS, 25 rue des Martyrs, 38042 Grenoble, France}
\author{Rahul R. Nair}
\author{A. K. Geim}
\affiliation{Department of Physics and Astronomy, University of
Manchester, Manchester, M13 9PL, United Kingdom}


\begin{abstract}
Local laser excitation and temperature readout from the intensity
ratio of Stokes to anti-Stokes Raman scattering signals are
employed to study the thermal properties of a large graphene
membrane. The concluded value of the heat conductivity coefficient
$\kappa\approx$~600~W/m$\cdot$K is smaller than previously
reported but still validates the conclusion that graphene is a
very good thermal conductor.
\end{abstract}

\maketitle


Systems of fewer than three dimensions are considered to be very
efficient heat spreaders as their intrinsic thermal conductivity
may eventually diverge for infinite specimens, following a power
or logarithmic law for one- or two-dimensional systems,
respectively~\cite{Lepri03}. This conjecture is now being
confronted with experimentations on graphene - a single sheet of
graphite, the closest archetype of a two-dimensional crystal and
promising material for various applications~\cite{Geim09}.
Although conventional methods to extract thermal properties of
solids are not easily applicable to a system of a single atomic
monolayer, the heat conductivity of graphene flakes has been shown
to be conveniently investigated using contactless methods of local
laser excitation combined with micro-Raman scattering
spectroscopy~\cite{Balandin08}. Carbon crystallites such as
diamond and graphite are known as the exceptional heat
conductors~\cite{Wei93} and an efficient thermal conductivity has
been also reported for graphene~\cite{Balandin08}. This conclusion
calls, however, for confirmation because the experimental methods
applied to draw it are not fully straightforward. A precise
(contactless) temperature readout, accurate sample geometry and
exact estimations of the absorbed laser power are among subtle
issues which may significantly influence the apparent values of
the extracted thermal conductivity coefficient.

In this paper we report on room temperature studies of thermal
properties of a relatively large graphene membrane~\cite{Nair08,Booth08}.
Our Corbino-like experimental configuration together with the
direct temperature readout from the intensity ratio of Stokes to
anti-Stokes Raman scattering signals largely simplifies the data
analysis. The presented results are in overall agreement with the
previous studies~\cite{Balandin08,Ghosh08} but we argue that the
extracted value of the 3D-equivalent thermal conductivity
coefficient for graphene may not be as high as it has been
reported so far.

A photograph of our sample, the graphene membrane, as seen through
a x100 microscope objective is presented in the inset of Fig. 1.
The membrane fully covers the 44 $\mu$m diameter
pinhole made in the 2 mm thick plate of copper. With the use of
silver epoxy, the edges of the membrane (which extend outside the
pinhole) are thermally short circuited to the copper plate which
serves in our experiments as a room temperature heat sink. The
suspended part of the membrane has a well defined circular
geometry.


\begin{figure}
\includegraphics[width=0.8\linewidth,angle=0,clip]{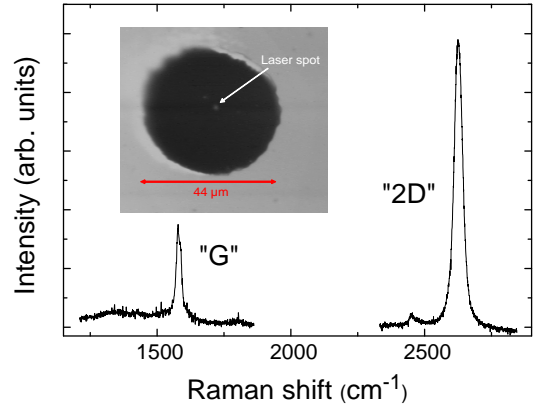}
\caption{\label{Fig1} The characteristic room temperature Raman scattering bands of
the investigated graphene membrane, measured under 6.2 mW
excitation of 632.8 nm He-Ne laser line focused on the middle of
the membrane. Inset : Optical photograph of the graphene membrane.
The diameter of the membrane is 44 $\mu$m and the white spot in
the middle is the diffusion of the laser spot of 2$\mu$m diameter
on the one atom thick graphene membrane.}
\end{figure}

The groundwork of our experiments consists in using laser
excitation to locally generate heat and measuring the Raman
scattering spectra to extract the actual temperature of the
membrane within the laser spot. The Raman scattering spectra shown in Fig. 1 reveal the
characteristic "G" and "2D" bands~\cite{Ferrari06,Graf07} of graphene,
measured on our membrane at room temperature when the laser spot
is located at its center.

The local temperature of the membrane within the laser spot is
derived from measurements of the Stokes and anti-Stokes Raman
scattering signals corresponding to the low energy ($\hbar \omega _{G}$) phonon
of graphene (G-band). After a careful calibration of the response of
the experimental set-up (with a black-body like tungsten radiation
source and a two-color pyrometer), we read this temperature directly
from the intensity ratio of the Stokes to anti-Stokes signals :
$\frac{I(\omega _{exc}-\omega _{G})}{I(\omega _{exc}+\omega _{G})}
= \exp(\frac{\hbar \omega _{G}}{k_{B}T})$ . Such simple method of
temperature readout is well justified when working with a single
atomic layer. Reabsorption processes, which usually need to be
taken into account for thick bulk samples, are negligible small in
our case. Few examples of the measured Stokes and anti-Stokes
components of the Raman scattering spectra of the G-band are shown
in Fig. 2. As expected and seen in this figure, the laser
excitation heats locally the membrane most efficiently when the
laser spot is at the center of the membrane and significantly less
when it is placed closer to the copper plate heat sink.

\begin{figure}
\includegraphics[width=0.9\linewidth,angle=0,clip]{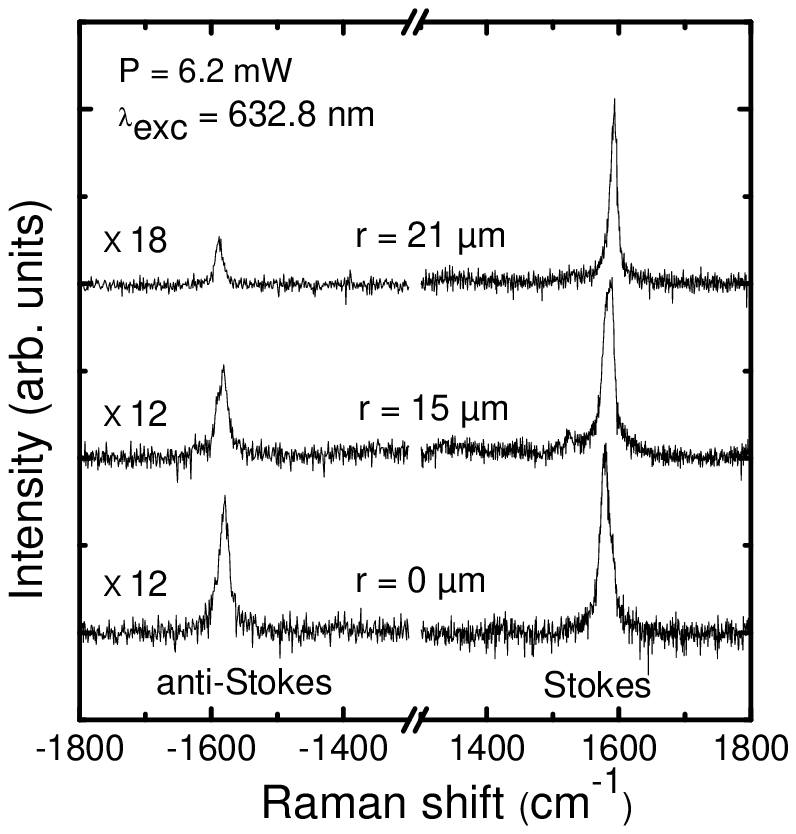}
\caption{\label{Fig2} Stokes and anti-Stokes Raman scattering
spectra of the the G-band graphene membrane measured under 6.2 mW
of laser excitation focused at different points on the membrane
(at different distances r from the center of the membrane). Note
the change in the ratio of Stokes to anti-Stokes signal, which
reflects the drop of local temperature within the laser spot
(2$\mu$m in diameter) when approaching the edge of the membrane
(in thermal contact with a room temperature sink). }
\end{figure}

As further described in details, the temperature difference
$\Delta T$ between the laser spot location and the membrane's edge
is proportional to the absorbed laser power $P$ and inversely
proportional to the efficiency of the "heat spread": $\Delta T
\sim P/\kappa d$, where $\kappa$ is the 3D equivalent thermal
conductivity coefficient and $d$ is the thickness of the membrane.
Knowing the proportionality factor, which depends on the geometry
of the experimental configuration and can be calculated from the
heat diffusion equation, we can extract the characteristic
parameters of heat conductivity in graphene.

In the calculations we consider that the heat generation
$q(\overrightarrow{r})$ in the membrane (disk of the radius $R$
and thickness $d$) is due to a local, homogenous across the membrane
but Gaussian-spread in plane, laser excitation:
$q(\overrightarrow{r})=\frac{P}{d \pi a^{2}
 }\exp(-|\overrightarrow{r}-\overrightarrow{r_{0}}|^2)/a^{2})$.
Here, $a = 1$~$\mu$m corresponds to the estimated radius of the
laser spot on the sample, $P =d\cdot\int
q(\overrightarrow{r})d^{2}\overrightarrow{r}$ is the total
absorbed laser power and $\overrightarrow{r}$ is the vector of the
in plane polar coordinates which we center in the middle of the
membrane. The temperature distribution in the membrane plane
$T(\overrightarrow{r})$, and in particular its measured value at
the location of the laser spot $T(\overrightarrow{r_{0}})$ is
ruled by the steady state form of the heat diffusion equation:
\begin{equation}
\kappa
\cdot\bigtriangledown^{2}T(\overrightarrow{r})+q(\overrightarrow{r})=0
\end{equation}
which after introducing the dimensionless variable
$\overrightarrow{\rho}=\overrightarrow{r}/a$ takes here the
following form :
\begin{equation}
\bigtriangledown^{2}T(\overrightarrow{\rho})+ \frac{P}{\kappa \pi
d}\exp(-|\overrightarrow{\rho}-\overrightarrow{\rho_{0}}|^2)=0
\label{equHeat}
\end{equation}

With fixed geometrical factors (R, a, $\overrightarrow{\rho
_{0}}$) and under appropriate boundary conditions (room
temperature at the edge of the membrane), the solution of equation~(\ref{equHeat})
depends on a single parameter $\alpha=P / \kappa d
\pi $. Conversely, the measure of temperature at any point of the
membrane, and in particular at the location of the laser spot,
allows us to determine $\alpha$, and also to extract the entire
temperature distribution in the membrane.

Equation (2) can be readily solved when the laser excitation
is focused at the center of the membrane. The solution of our heat
diffusion equation has a circular symmetry
($T(\overrightarrow{r})=T(r)$) and takes the following form :

$T(0)-T(\rho)=\frac{\alpha}{2} \int _{0}^{\rho}
\frac{1-\exp(-x^{2})}{x} dx =$

$ = \frac{\alpha}{2}(\ln \rho -\frac{1}{2}Ei(-\rho^{2}) +
\frac{\gamma}{2} ) \approx \frac{\alpha}{2}(\ln \rho
+\frac{\gamma}{2})$

where Ei(x) denotes the exponential integral function, $\gamma =
0.5772$ is the Euler's constant and the approximation is
well satisfied if $\rho > 1$.


In the experiment, the edges of the membrane are kept at ambient
condition $T(\Lambda)=T_{edge}$ = 295~K and $\Lambda$=R/a~=~22. Thus
$T(0)-T_{edge} \cong 1.689 \alpha $.
Under a 6.2 mW laser excitation, we obtain T(0)=660 K at the laser spot
in the middle of the membrane and immediately obtain $\alpha$~=~216~K.

The parameter $\alpha$ can also be extracted for an arbitrary experimental geometry
(laser spot out of the center of the membrane) but then numerical solutions of the heat flow equation
are required. This has been done using finite elements
computations. As shown in Fig. 3, the measured local
temperature within the laser spot located now at different
positions with respect to the center of the membrane is well
reproduced by the simulations assuming $\alpha$~=~214.4~K. This
value is not far from the one derived previously from a single measurement
with the laser placed at the center of the membrane.

\begin{figure}
\includegraphics[width=0.6\linewidth,angle=0,clip]{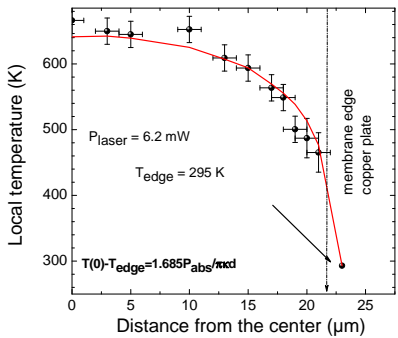}
\caption{\label{Fig3} Lattice temperature as deduced from the
intensity ratio of Stokes and anti-Stokes signals of the graphene
G-band raman scattering signals, (black dots), measured at room
temperature with a laser power of 6.2 mW focused down to 2 $\mu$m
diameter spot. The data point on the copper plate is fixed at
T=295~K. The solid red line is the solution of the temperature
profile obtained by finite elements simulation with $\alpha$ =
214.4 K.}
\end{figure}

After evaluating $\alpha$ we can now extract the thermal
conductivity coefficients of graphene but this requires an
estimation of both the fraction of the laser power absorbed by the
sample and of the thickness of the membrane.
In our analysis, we assume a single laser pass through the membrane, which is
justified in our experiment because the substrate below the membrane
has been completely removed, preventing any parasitic back reflections. It has been demonstrated and
confirmed by different studies~\cite{Kuzmenko08,Nair08,Mak08} that the absorption of light in the visible
range of energy in a graphene monolayer is determined only by fundamental constants,
and is equal to $\pi e^2/(\hbar c 4 \pi \epsilon_0)=\pi\alpha\approx2.3\%$, where $\alpha$ is the fine structure constant. It follows from these studies that 2.3\% of the total
laser power (6.2 mW) is absorbed by the membrane. Hence, if $\alpha$~=~214.4~K,
the "two-dimensional" thermal coefficient of graphene is $\kappa \cdot d = 2.117 \cdot
10^{-7}$~W/K. Assuming $d~=~$0.335~nm, corresponding to the inter
layer distance in graphite, a 3D equivalent value $\kappa$=632~W/m$\cdot$K
is obtained.

This value of $\kappa$ is by a factor 5 to 8 smaller
than the one reported previously~\cite{Balandin08} but nevertheless indicates that
heat spreading in graphene is as efficient as it is in
graphite~\cite{Slack62}, the latter considered as an exceptional
heat conductor. The difference between the present and previous
estimations of $\alpha$ for graphene is mainly due to different
assumptions regarding the absorbance of graphene. Our assumption
of 2.3\% of the absorbed laser power follows the results of very
recent and precise transmission and reflectivity studies of graphene (membranes or deposited on a substrate) and graphite~\cite{Kuzmenko08,Nair08,Mak08}, and
seems to be more realistic as compared to 13\% assumed
previously~\cite{Balandin08}. (Supposing 13\% of the absorbed
laser power, we would conclude $\kappa\approx$~3600~W/m$\cdot$K,
in fair agreement with previous data). We also note that our
direct readout of local temperature confirms the applicability of
the method used by Balandin \textit{et al.}~\cite{Balandin08} to
measure this temperature, which relies on the power dependent shift of
the G-band, which on its hand is attributed to a temperature dependent shift
known from independent experiments. When exciting at the middle of
our membrane with different laser powers, we find that the G band Raman shift
follows a linear variation with increasing optical power $\omega_{G}(P) = \omega _{G}(0) - P\cdot0.8 cm^{-1}/mW$,
where $\omega _{G}(0)=1589.7 cm^{-1}$ is the limit for a vanishing excitation power.
Assuming that $\Delta \omega/\Delta T = -0.0016 cm^{-1}/K$~\cite{Calizo07},
we conclude that an increase of 1 mW of the excitation power corresponds to
an increase of 50 K of the lattice temperature. Thus at the 6.2~mW excitation:
$T~=~$605~K which is not far from $T~=~$660~K measured from
the intensity ratio of Stokes to anti-Stokes signals.

To conclude, we have used micro-Raman scattering experiments to
study the room temperature heat conductivity of a large graphene
membrane. We have deduced that graphene is a thermal conductor as good as
graphite. The 3D equivalent thermal coefficient of graphene is
$\kappa\approx$~630~W/m$\cdot$K, i.e., somewhat smaller than the values
previously reported~\cite{Balandin08}. The difference between the
present and previous estimations of $\kappa$ is mainly due to
different assumptions regarding the efficiency of the
graphene's optical absorbance.

\section{Methods}

Free standing graphene membranes were prepared by the method previously reported\cite{Booth08}. In brief, large (>> 100 $\mu$m in size) graphene crystals were deposited on top of a silicon wafer, which was spin coated with a 90nm thick layer of PMMA. Single layers were identified by optical microscopy. By employing a series of photolithography and electro-deposition steps, we deposited a 15 to 20 $\mu$m thick copper film on top of the wafer. The film contained an opening of 50 to 100 $\mu$m in diameter, which was aligned with the chosen graphene crystal so that graphene fully covered the aperture. We used acetone to dissolve PMMA and thus release the copper scaffold with graphene attached into the liquid. The samples were finally dried in a critical pint dryer to prevent the membrane rupturing due to surface tension. Silver epoxy was used to place the scaffold to a thick copper plate to allow easy handling and the reported measurements.

Raman scattering experiments have been carried out at room temperature using the 632.8 nm line of the He-Ne laser
as the excitation source and a confocal micro-Raman set-up
equipped with a x100 microscope objective which provides a lateral
resolution of $\sim$2$\mu$m (diameter of the laser spot on the
membrane). An X-Y translation stage together with an imaging
camera allowed us to place the laser spot on the membrane at a
given location with a precision of 0.1 $\mu$m; the excitation power
is measured at the sample location with a calibrated silicon
photodiode.

\begin{acknowledgements}
We gratefully acknowledge A-M. Bonnot and E. Bustarret for
stimulating discussions. Part of this work was supported by ANR
project ANR-08-JCJC-0034-01, by EC Grant
MTKD-CT-2005-029671, by PCR CNRS-NRC and by GACR No.~P204/10/1020.
\end{acknowledgements}

\bigskip

\providecommand*{\mcitethebibliography}{\thebibliography}
\csname @ifundefined\endcsname{endmcitethebibliography}
{\let\endmcitethebibliography\endthebibliography}{}

\end{document}